\documentclass[%
  reprint,
  superscriptaddress,
  amsmath,amssymb,
  aps,
  prl,
  longbibliography,
  floatfix,
]{revtex4-2}

\usepackage{graphicx}
\usepackage{dcolumn}
\usepackage{bm}
\usepackage{xcolor}
\usepackage[hidelinks]{hyperref}
\graphicspath{{figures/}}

\begin{document}


\title{Generalized Theory of Blowout-Regime Plasma Wakefields via a Lagrangian Formulation}

\author{Y. Kang}
\thanks{conrad278@g.ucla.edu}
\affiliation{Department of Physics and Astronomy, UCLA, Los Angeles, California 90095, USA}
\author{P. Manwani}
\affiliation{Department of Physics and Astronomy, UCLA, Los Angeles, California 90095, USA}
\author{S. S. Baturin}
\affiliation{School of Physics and Engineering, ITMO University, 197101 St. Petersburg, Russia}
\author{G. Andonian}
\affiliation{Department of Physics and Astronomy, UCLA, Los Angeles, California 90095, USA}
\author{B. Naranjo}
\affiliation{Department of Physics and Astronomy, UCLA, Los Angeles, California 90095, USA}
\author{J. B. Rosenzweig}
\affiliation{Department of Physics and Astronomy, UCLA, Los Angeles, California 90095, USA}


\date{\today}

\begin{abstract}
The blowout regime of plasma wakefield acceleration is a strongly nonlinear plasma response in which an intense relativistic beam or laser pulse expels plasma electrons forming a nearly uniform ion column. Existing self-consistent theories describe this structure primarily in the axisymmetric limit, reducing the bubble dynamics to a single radial degree of freedom. This restriction is inadequate for flat beams with highly asymmetric transverse emittances, as required in future linear colliders to control luminosity and beamstrahlung. Using a reduced Lagrangian formulation, we generalize the existing theory beyond axisymmetry and derive a coupled system of equations governing the wake potential driven by an asymmetric beam. This theory self-consistently predicts the focusing fields and their variation along the co-moving coordinate, as well as the accelerating field and its transverse nonuniformity. The resulting theoretical predictions are benchmarked against particle-in-cell simulations, demonstrating the theory's ability to describe and interpret wakefields driven by nonaxisymmetric sources.

\end{abstract}

\maketitle

Plasma  acceleration uses an intense laser pulse or a relativistic charged-particle bunch to drive a plasma wake capable of sustaining accelerating gradients far beyond those of conventional radio-frequency structures~\cite{Tajima1979,Chen1985}. In this paper, we are concerned with the case of an electron beam driver, a scenario referred to as plasma wakefield acceleration (PWFA). Modern PWFA research is concentrated on the blowout regime, where a sufficiently intense driver expels nearly all plasma electrons from a finite near-axis region, leaving an ion cavity bounded by a thin sheath of expelled electrons~\cite{Rosenzweig1991,Pukhov2002,Kostyukov2004,Lotov2004}. The evacuated cavity provides linear transverse focusing from the exposed ion column and an accelerating field that is, to leading order, independent of transverse offset near the cavity axis, enabling high-gradient acceleration while preserving beam quality. The corresponding axisymmetric theory of the blowout regime is now well developed. Lu et al.\, reduced the quasi-static plasma response to a closed second-order equation for the evolution of the cavity radius \(r_b\), from which the electromagnetic fields follow~\cite{Lu2006}. Subsequent investigations extended this work to nonlinear beam loading and increasingly detailed descriptions of the electron sheath and return-current structure~\cite{Tzoufras2008,Yi2013,Dalichaouch2021,Liu2025}. In a complementary approach, Golovanov et al.\, used the conservation of the wake quasi-energy to derive a modified equation for the cavity boundary  \(r_b\)~\cite{Golovanov2023}.

The need to move beyond an axisymmetric description arises directly from application of plasma acceleration for future \(e^{+}e^{-}\) linear colliders. Designs for these discovery machines use \textit{flat }beams which possess a large horizontal-to-vertical normalized emittance ratio, \(\epsilon_{n,x}/\epsilon_{n,y}\), to obtain high luminosity while controlling the damaging effects of beamstrahlung at the interaction point~\cite{Chen1992,Adolphsen2013,Aicheler2012}. Plasma-based collider concepts would therefore have to preserve, transport, and focus strongly asymmetric beams~\cite{Chen1990,Foster_2023_HALHF}. Flat driving beams produce nonaxisymmetric blowout cavities, with unequal focusing forces and transverse structure in the longitudinal accelerating field~\cite{Baturin2022,Manwani2025, Kang2024}. It is essential  to be able to self-consistently predict these fields for matching a flat witness beam through an accelerating stage ~\cite{manwani_2022_ipac} or for designing plasma lens systems focusing flat beams. This requirement is especially important for preserving the transverse emittance ratio while accelerating flat beams, since any nonlinear transverse coupling can resonantly exchange the horizontal and vertical emittances~\cite{Diederichs2024}. 
Generally, longitudinal variation of the asymmetric focusing profiles is accompanied, according to the Panofsky--Wenzel theorem~\cite{Panofsky1956}, by transverse dependence of \(E_z\). In addition, broken axisymmetry also permits \(B_z\) and higher-order focusing non-uniformities. Without axisymmetry, the blowout cavity boundary and the wake potential associated with it carry multiple independent transverse degrees of freedom, and neither a single trajectory equation~\cite{Lu2006} nor a single scalar conservation law~\cite{Golovanov2023} is sufficient to close their coupled dynamics. 
Because of the compelling applications in particle physics discovery accelerators, interest in nonaxisymmetric plasma wakes has increased in recent years, with significant  theoretical work undertaken~\cite{Baturin2022,Zhou2021,Berceanu2026}. In particular, an elliptical-cavity model has been used to derive the wake potential for an elliptical blowout, showing that the transverse fields remain linear in $x$ and $y$. Further, an analytical equilibrium estimate for the cavity boundary in the long-beam limit, \(\sigma_z\gg1\)~\cite{Manwani2025}, has been obtained. However, a self-consistent theory that predicts the longitudinal evolution of the cavity asymmetry for general beam distributions, together with the associated accelerating-field structure, has remained elusive so far.

Within this context, we introduce in this Letter a Lagrangian formulation of the quasi-static blowout system as a new route to a versatile and highly effective generalized theory of wake dynamics. Representing the transverse wake potential structure by a finite set of generalized coordinates, the resulting Euler--Lagrange equations provide the required closure as a coupled system of ordinary differential equations. The theory recovers the energy-conserving axisymmetric boundary equation of Ref.~\cite{Golovanov2023} and extends the self-consistent prediction of the blowout geometry and electromagnetic wakefields beyond the axisymmetric limit.

In this work, we employ dimensionless variables, where the number densities are normalized to the unperturbed plasma density \(n_0\), which defines the plasma frequency \(\omega_p=(n_0 e^2/\varepsilon_0 m_e)^{1/2}\) in SI units. Time is then normalized to \(\omega_p^{-1}\), lengths to \(k_p^{-1}=c/\omega_p\), charges to \(e\), charge densities to \(en_0\), velocities to \(c\), momenta to \(m_e c\), electric fields to \(E_0=m_e c\omega_p/e\), magnetic fields to \(B_0=E_0/c\), the scalar potential to \(m_e c^2/e\), and the vector potential to \(m_e c/e\). Here, \(e\) is the  elementary charge, \(m_e\) is the electron mass, \(c\) is the speed of light, and \(\varepsilon_0\) is the vacuum permittivity. The subscripts \(i\), \(e\), and \(b\) denote plasma ions, electrons, and the drive beam, respectively. For a driver propagating in the positive \(z\) direction, we introduce the co-moving variables \((x,y,z,t)\rightarrow(x,y,\xi,s)\), with \(\xi=t-z\) and \(s=z\). We assume that the driver evolves slowly compared with the plasma response in the co-moving frame and employ the quasi-static approximation \(\partial_s\ll \partial_\xi\) for all plasma quantities~\cite{Sprangle1990,Mora1997}. The plasma is taken to be cold and initially unperturbed, so electrons ahead of the driver carry zero velocity.  The ions are approximated as  immobile, as \(m_i\gg m_e\).

The Lagrangian formulation of collisionless plasma dynamics was initially introduced by Low~\cite{Low1958} and later applied to plasma wakefield theory~\cite{Chen1993,Brizard1998}. We begin our treatment by writing, under the assumptions given above, the quasi-static Lagrangian density
\begin{equation}
\begin{aligned}
\mathcal{L}_{\mathrm{QS}}
&=\frac{1}{2}
\left(
\mathbf{E}^2 - \mathbf{B}^2
\right)-n_i\phi-\rho_b\psi
\\
&+\sum_\alpha n_{e,\alpha}
\left(
1-\gamma_{e, \alpha}^{-1}+\phi-\mathbf{v}_{e, \alpha}\cdot\mathbf{A}
\right).
\end{aligned}
\label{eq:LQS-main}
\end{equation}
The homogeneous Maxwell equations are enforced by expressing the electromagnetic fields in terms of potentials, \(\mathbf{E}=-\nabla\phi-\partial_t\mathbf{A}\) and \(\mathbf{B}=\nabla\times\mathbf{A}\). The introduction of the wake potential, \(\psi=\phi-A_z\), permits substitution of \(A_z\) with \(\psi\), so that the Lagrangian density can be written compactly as \(\mathcal{L}_{\mathrm{QS}}=\mathcal{L}_{\mathrm{QS}}(\phi,\psi,\mathbf{A}_\perp;x,y,\xi)\). The index \(\alpha\) labels electron streams, allowing a description of trajectory crossing in the nonlinear regime. Within each stream, the momentum is single-valued at a given position, consistent with the cold-fluid description. We may therefore define the total electron density \(n_e=\sum_\alpha n_{e,\alpha}\) and the density-weighted electron energy and velocity as \(\bar{\gamma}_e = n_e^{-1}\sum_\alpha n_{e,\alpha}\gamma_{e,\alpha}\) and \(\bar{\mathbf{v}}_e = n_e^{-1}\sum_\alpha n_{e,\alpha}\mathbf{v}_{e,\alpha}\). As the ions are considered immobile, the ion and driver-beam terms have only a source--field interaction Lagrangian contribution. The ultra-relativistic driver distribution is quasi-statically prescribed as \(J_{b,z}\simeq\rho_b\) so it couples to the plasma response only through \(\psi\). Integrating Eq.~\eqref{eq:LQS-main} in the transverse plane gives \(L_{\mathrm{QS}}=\int d^2x_\perp\,\mathcal{L}_{\mathrm{QS}}=L_{\mathrm{EM}}+L_{\mathrm{electron}}+L_{\mathrm{ion}}+L_{\mathrm{driver}}\). The quasi-static action is therefore \(\mathcal{S}_{\mathrm{QS}}=\int d\xi\,L_{\mathrm{QS}}\), defined as the action per unit length in the longitudinal coordinate \(s\). Varying \(\mathcal{L}_{\mathrm{QS}}\) with respect to \(\phi\) and \(\mathbf{A}_\perp\) gives the quasi-static field relations
\begin{subequations}
\label{eq:qs-maxwell-identities}
\begin{align}
\nabla_{\perp}^{2}\psi &= -1 + n_e (1-\bar{v}_z),
\label{eq:qs-poisson-identity}\\[1ex]
\mathbf J_\perp &= \nabla_\perp\partial_\xi\psi - \hat{\mathbf z}\times\nabla_\perp B_z .
\label{eq:qs-current-identity}
\end{align}
\end{subequations}
where \(n_i=1\) has been used, and \(\mathbf J_\perp\equiv -n_e\bar{\mathbf v}_\perp\) is the transverse electron current density. These are the inhomogeneous Maxwell equations in quasi-static form. 

Equation~\eqref{eq:qs-poisson-identity} is the Poisson equation for the wake potential \(\nabla_\perp^2\psi=S\) with the conventional wake source \(S=-\left(\rho-J_z\right)\), following from Gauss's law and the longitudinal component of Maxwell--Amp\`ere's law. Integrating the continuity equation over the transverse plane shows that the integral of the source is conserved for every transverse slice, \(\frac{d}{d\xi}\int d^2x_\perp S = 0\), assuming vanishing transverse flux at infinity. Equation~\eqref{eq:qs-current-identity} is the transverse component of Maxwell--Amp\`ere's law. 

\begin{figure}
    \centering
    \includegraphics[width=\columnwidth]{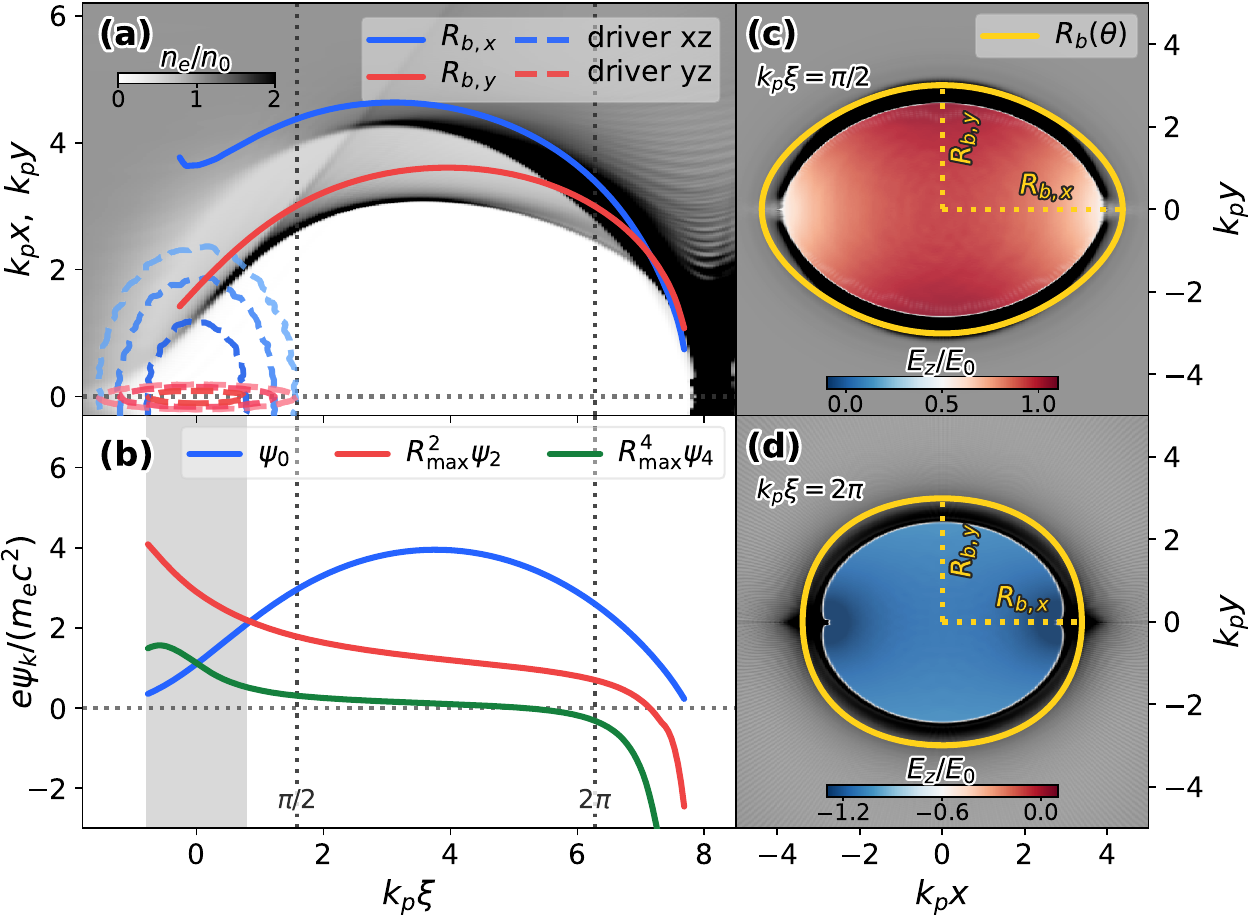}
    \caption{PIC simulation of an asymmetric Gaussian driver centered at \(\xi=0\), with peak density \(n_b=32\), rms length \(\sigma_z=0.8\), transverse aspect ratio \(\sigma_x/\sigma_y=12\), and \(\sigma_x\sigma_y=0.125\). (a) Electron density \(n_e/n_0\) in overlaid \(x\)-\(\xi\) and \(y\)-\(\xi\) slices, with the simulation-inferred effective boundary along the two axes, \(R_{b,x}=R_b(0)\) and \(R_{b,y}=R_b(\pi/2)\) (solid), and driver density contours (dashed). (b) Wake moments \(\psi_k\) for \(k=0,2,4\), obtained by projecting the simulated wake potential onto Eq.~\eqref{eq:even_multipole_expansion}; each is scaled by \(R_{\max}^{k}\), with \(R_{\max}\) the maximum blowout radius, so that all curves share the units of potential. The shaded band marks the driver region, \(|\xi|\le\sigma_z\). (c),(d) Transverse slices of \(E_z/E_0\) at \(\xi=\pi/2\) and \(2\pi\) (vertical dotted lines in (a) and (b)), overlaid on the electron density, with the inferred effective boundary \(R_b(\theta)\) in yellow.}
    \label{fig:asymmetry}
\end{figure}

The quasi-static Lagrangian formulation becomes a predictive theory only after imposing the geometric structure specific to the blowout regime described above: an ion column enclosed by a thin sheath of  plasma electrons expelled by the fields of the intense beam. The maximum excursion from the axis of the sheath permits classification of the blowout; if this position in both $x$ and $y$ greatly exceeds unity, one describes this as a strong blowout. Following simulation observations and previous theory~\cite{Golovanov2023,Manwani2025}, we adopt a model that idealizes the electron sheath as a $\delta$-function layer representing the essential effects of the nonlinear plasma electron distribution dynamics. That is, instead of being a physical description of the complicated electron sheath profile, this equivalent $\delta$-sheath serves to produce the electromagnetic fields\textit{ inside} the blowout cavity accurately, without need of additional details. The equivalent $\delta$-sheath boundary that generates similar electromagnetic fields as the physical finite-thickness sheath is expected to lie slightly outside the actual boundary between the ion column and the electron sheath, as seen in Fig.~\ref{fig:asymmetry}. The difference between the $\delta$-sheath position and the actual cavity boundary is negligible in the strong-blowout regime, but becomes increasingly noticeable for weaker blowout. In the weak blowout limit, a larger and more convex boundary, \(r_b\) in the case of an axisymmetric system, is needed to produce the correct slope of the longitudinal field near the blowout center, where
\(E_z' \simeq r_b r_b''/2\). Further simulation verification of the equivalent $\delta$-sheath geometry, together with the corresponding multipole expansion of the wake potential, is presented in Ref.~\cite{Yunbo_AAC_Paper_2026}.

\begin{figure}
    \centering
    \includegraphics[width=\columnwidth]{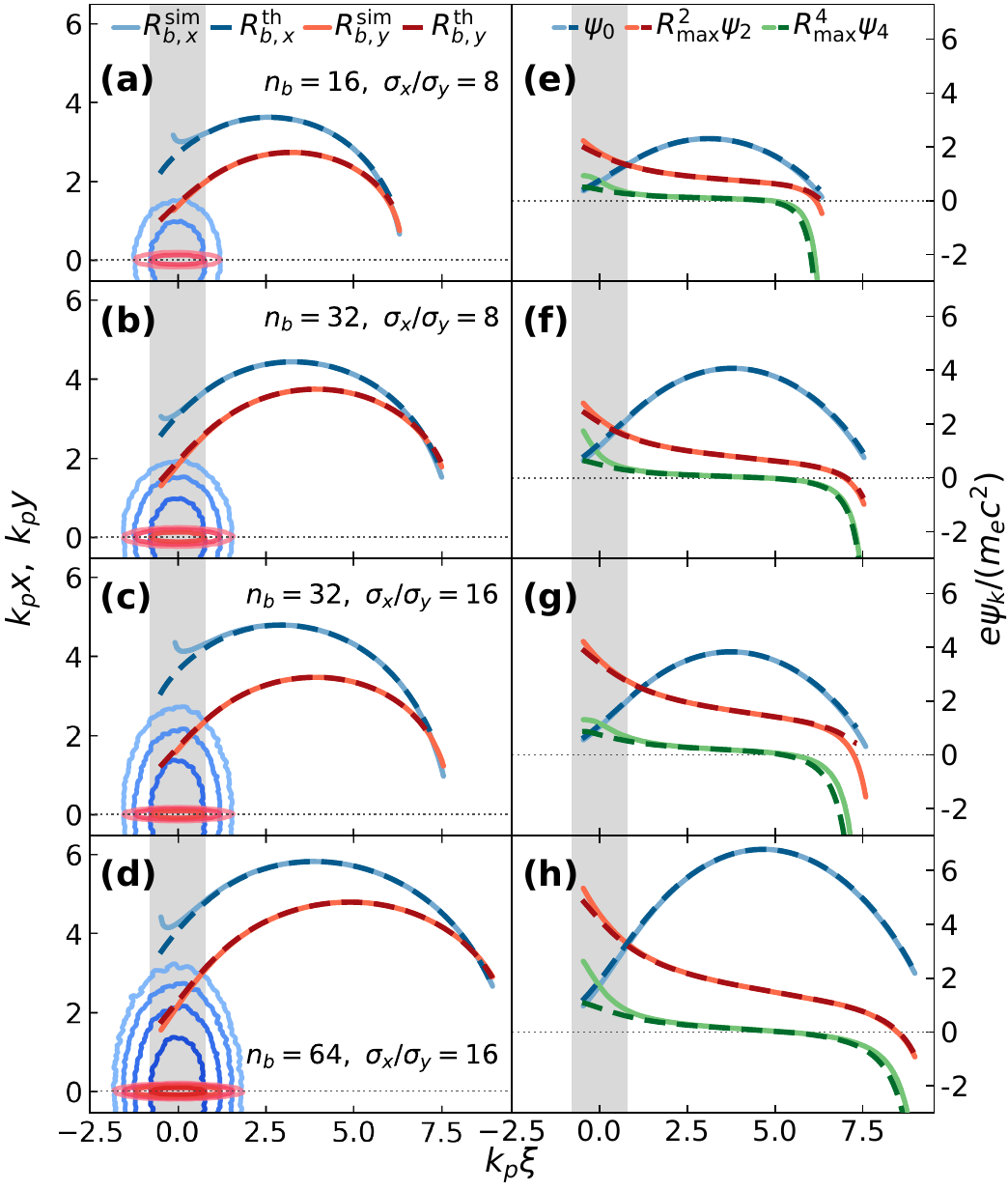}
    \caption{Comparison between theory (dashed) and PIC simulation (solid), following the conventions of Fig.~\ref{fig:asymmetry}, for four asymmetric drivers with the same \(\sigma_x\sigma_y=0.125\) and \(\sigma_z=0.8\), with \((n_b,\sigma_x/\sigma_y)\) annotated in each row. (a)--(d) Effective boundary radii \(R_{b,x}\) and \(R_{b,y}\), and (e)--(h) the corresponding wake moments \(\psi_0\), \(\psi_2\), and \(\psi_4\). Theoretical curves are obtained by integrating Eq.~\eqref{eq:general_moment_equations}.}
    \label{fig:moments}
\end{figure}

To formulate a general $\delta$-sheath model, we consider a transverse slice containing the two-dimensional blowout region \(\Omega_b\), filled with the uniform ion background. Its boundary is denoted by \(\partial\Omega_b\), and the
plasma electrons are collapsed onto it as a surface source density \(\sigma_S\), defined by \(\sigma_S\,d\Gamma=n_e(1-\bar v_z)\,d^2x_\perp\), where \(d\Gamma\) indicates a boundary line element. Because in our model the $\delta$-sheath subsumes the entire plasma response, the plasma outside remains unperturbed and the wake fields are strictly shielded beyond the cavity. Equations~\eqref{eq:qs-maxwell-identities} then reduce to
\begin{subequations}
\label{eq:identities_delta}
\begin{align}
&\nabla_{\perp}^{2}\psi
= S = -\mathbf{1}_{\Omega_b} + \sigma_S\delta_{\partial\Omega_b},
\qquad
\sigma_S = -\partial_n\psi_{\rm in},
\label{eq:identity-delta-poisson}\\[0.3cm]
&\nabla_\perp B_z
= -\hat{\mathbf z}\times\nabla_\perp\partial_\xi\psi,
\qquad
J_\tau = \left.B_z\right|_{\partial\Omega_b}.
\label{eq:identity-delta-current}
\end{align}
\end{subequations}
Here, \(\mathbf{1}_{\Omega_b}\) equals unity inside \(\Omega_b\) and vanishes outside. We henceforth denote \(n\) and \(\tau\) as the normal and tangential directions on \(\partial\Omega_b\), respectively, with \(\hat{\mathbf n}\times\hat{\boldsymbol{\tau}}=\hat{\mathbf z}\), so that \(\mathbf{\hat{n}}\cdot \nabla_\perp = \partial_n\) and \(J_\tau = \hat{\boldsymbol{\tau}}\cdot\mathbf{J}_\perp\). The relations for \(\sigma_S\) and \(J_\tau\) follow from integrating Eqs.~\eqref{eq:identities_delta} across the sheath and using the exterior shielding condition. Inside the cavity, Equation~\eqref{eq:identity-delta-current} states the Cauchy--Riemann relation, closing \(B_z\) in terms of wake potential. For a centered quasi-static drive beam with even parity in both transverse coordinates, the wake potential inherits the symmetry
\(\psi(x,y,\xi)=\psi(-x,y,\xi)=\psi(x,-y,\xi)\).  This permits an even multipole
expansion of the interior solution of Eq.~\eqref{eq:identity-delta-poisson},
\begin{equation}
\psi(r,\theta;\xi)
=
-\frac{r^2}{4}
+
\sum_{k=0,2,4,\ldots}^{\infty}
\psi_{k}(\xi)\,r^{k}\cos(k\theta),
\label{eq:even_multipole_expansion}
\end{equation}
where the multipole expansion coefficients of the wake potential,
\(\psi_k(\xi)\), are hereafter referred to as the \textit{wake moments}.

Continuity of \(\psi\) across the sheath in combination with exterior shielding implies \(\psi|_{\partial\Omega_b}=0\), which in turn defines the effective blowout boundary \(R_b(\theta;\{\psi_k\})\) implicitly through the wake moments. Retaining only the monopole moment, \(k=0\), gives the axisymmetric case, \(R_b=r_b(\xi)\) and \(\psi_{\rm axi}=r_b^2/4-r^2/4\)~\cite{Lu2006,Golovanov2023}. In the simplest asymmetric blowout example, adding the quadrupole moment, \(k=2\), gives the elliptical boundary assumed in Ref.~\cite{Manwani2025},
parameterized by the semi-axes \(a_p\) and \(b_p\). Higher moments describe further departures from this elliptical approximation. In Fig.~\ref{fig:asymmetry}, the \(k=4\)  octupole correction produces a visible deformation of the effective boundary, indicating a higher-order structure in the blowout profile. This motivates formulating the blowout dynamics with the wake moments as generalized coordinates, rather than the geometric variables that are naturally suited only for round or elliptical cavities. 

Inside \(\Omega_b\), the wake fields follow directly from Eq.~\eqref{eq:even_multipole_expansion}: the transverse  fields
\begin{equation}
\mathbf{W}_\perp
=
-\nabla_\perp\psi
=
\frac{1}{2}\mathbf{r}_\perp
-
\sum_{k=2,4,\ldots}
\psi_k\,\nabla_\perp
\left[r^k\cos(k\theta)\right],
\label{eq:focusing_field}
\end{equation}
and the longitudinal fields, 
\begin{equation}
\begin{aligned}
E_z
&=
\partial_\xi\psi
=
\sum_{k=0,2,4,\ldots}
\psi_k' r^k\cos(k\theta), 
\\
B_z
&=
-\sum_{k=2,4,\ldots}
\psi_k' r^k\sin(k\theta),
\end{aligned}
\label{eq:longitudinal_field}
\end{equation}
where \(B_z\) is obtained from the Cauchy--Riemann closure Eq.~\eqref{eq:identity-delta-current}, and the prime henceforth denotes \(d/d\xi\).

Furthermore, within the $\delta$-sheath description, we characterize the collective motion of the sheath electrons, which are confined to move along the blowout surface, by the density-weighted average electron velocity \(\bar{\mathbf v}_e\)~\cite{Golovanov2023}. This relates the boundary longitudinal derivative to the averaged electron velocities \(\mathbf{r}'_\perp = \frac{1}{1-\bar{v}_z}\bar{\mathbf{v}}_\perp = V_n \hat{\mathbf{n}} + V_{\tau}\hat{\boldsymbol{\tau}}\). Differentiating the condition \(\psi|_{\partial\Omega_b}=0\) with respect to \(\xi\) gives \(V_n=E_z|_{\partial\Omega_b}/\sigma_S\), while the collapsed tangential current density, \(J_\tau=-\sigma_S V_\tau\), together with Eq.~\eqref{eq:identity-delta-current}, gives \(V_\tau=-B_z|_{\partial\Omega_b}/\sigma_S\). The well-known conservation of \(H - P_z\), where \(H_e = \bar{\gamma}_e - \phi\) is the plasma electron Hamiltonian and the associated canonical momentum, \(P_{e,z} = \bar{p}_{e,z} - A_z\), gives the familiar identity for an initially cold plasma, \(\bar{\gamma}_e - \psi - \bar{p}_{e,z} = 1\)~\cite{Mora1997}. Using these identities and \(\psi|_{\partial\Omega_b}=0\), the electron Lagrangian per unit sheath source reduces to \(\mathcal L_S =\frac12|\mathbf r'_\perp|^2+\phi-\mathbf r'_\perp\cdot\mathbf A_\perp\), so that \(L_i+L_e=-\int_{\Omega_b}d^2x_\perp\,\phi +\oint_{\partial\Omega_b}\sigma_S\,\mathcal L_S\,d\Gamma\). Invoking Eq.~\eqref{eq:identities_delta} again, we eliminate the explicit dependence on \(\phi\) and \(\mathbf A_\perp\) and obtain a reduced Lagrangian that depends only on the wake potential \(\psi\),

\begin{equation}
\begin{aligned}
L_{\rm red} &=\frac{1}{2}\int_{\Omega_b}d^2x_\perp \left[ E_z^2+B_z^2 -\psi \right]\\
&+\frac{1}{2}\oint_{\partial\Omega_b}(V_n^2+V_\tau^2)\sigma_S d\Gamma - \int_{\Omega_b}d^2x_\perp\rho_b\psi,
\end{aligned}
\label{eq:reduced-lagrangian}
\end{equation}
where we choose the wake moments, \(\psi_{k}(\xi)\), as the generalized coordinates for the Euler-Lagrange formulation. We denote the terms quadratic in \(\psi_k'\) by \(K\), the terms independent of \(\xi\)-derivatives by \(U\), and the driver term by \(D\), so that \(L_{\mathrm{red}}=K-U-D\). The Euler--Lagrange equations \(\frac{d}{d\xi}(\partial L_{\rm red}/\partial\psi_k') -\partial L_{\rm red}/\partial\psi_k=0\) take the form
\begin{equation}
\begin{aligned}
\sum_{\ell }
A_{k\ell}\psi_\ell''
+
\sum_{\ell,m}
B_{k\ell m}\psi_\ell'\psi_m'
+
C_k
&=
\lambda_k,
\end{aligned}
\label{eq:general_moment_equations}
\end{equation}
with the coefficients 
\(A_{k\ell} = \int_0^{2\pi}d\theta\bigl[\frac{R_b^{k+\ell+2}(\theta)}{k+\ell+2}-\frac{R_b^{k+\ell+1}(\theta)}{\partial_r\psi(R_b,\theta)}\bigr]\cos[(k-\ell)\theta]\), \(B_{k\ell m} = \frac12(\partial_{\psi_\ell}A_{km}+\partial_{\psi_m}A_{k\ell}-\partial_{\psi_k}A_{\ell m})\), \(C_k = \frac{1}{2(k+2)}\int_0^{2\pi}d\theta R_b^{k+2}(\theta)\cos(k\theta)\), and \(\lambda_k = -\int_0^{2\pi}d\theta\int_0^{R_b(\theta)}dr\rho_b(\xi,r,\theta)r^{k+1}\cos(k\theta)\) for \(k,\ell,m=0,2,4,\ldots\). The detailed derivations of Eq.~\eqref{eq:reduced-lagrangian} and Eq.~\eqref{eq:general_moment_equations} are provided in the Appendix. In the axisymmetric limit, we take $k= 0$ and only the monopole term survives. Equation~\eqref{eq:general_moment_equations} reduces to the same axisymmetric ODE in Ref.~\cite{Golovanov2023}. The energy balance equation is recovered here from the canonical-energy evolution implied by \(L_{\rm red}\), \(d\mathcal{H}/d\xi=-\partial_\xi L_{\rm red}\) and \(\mathcal{H} = K+U +D\). The Lagrangian, however, contains richer information: it governs the coupled evolution of all the wake moments, as required to describe the dynamics of asymmetric blowouts.

\begin{figure}
    \centering
    \includegraphics[width=\columnwidth]{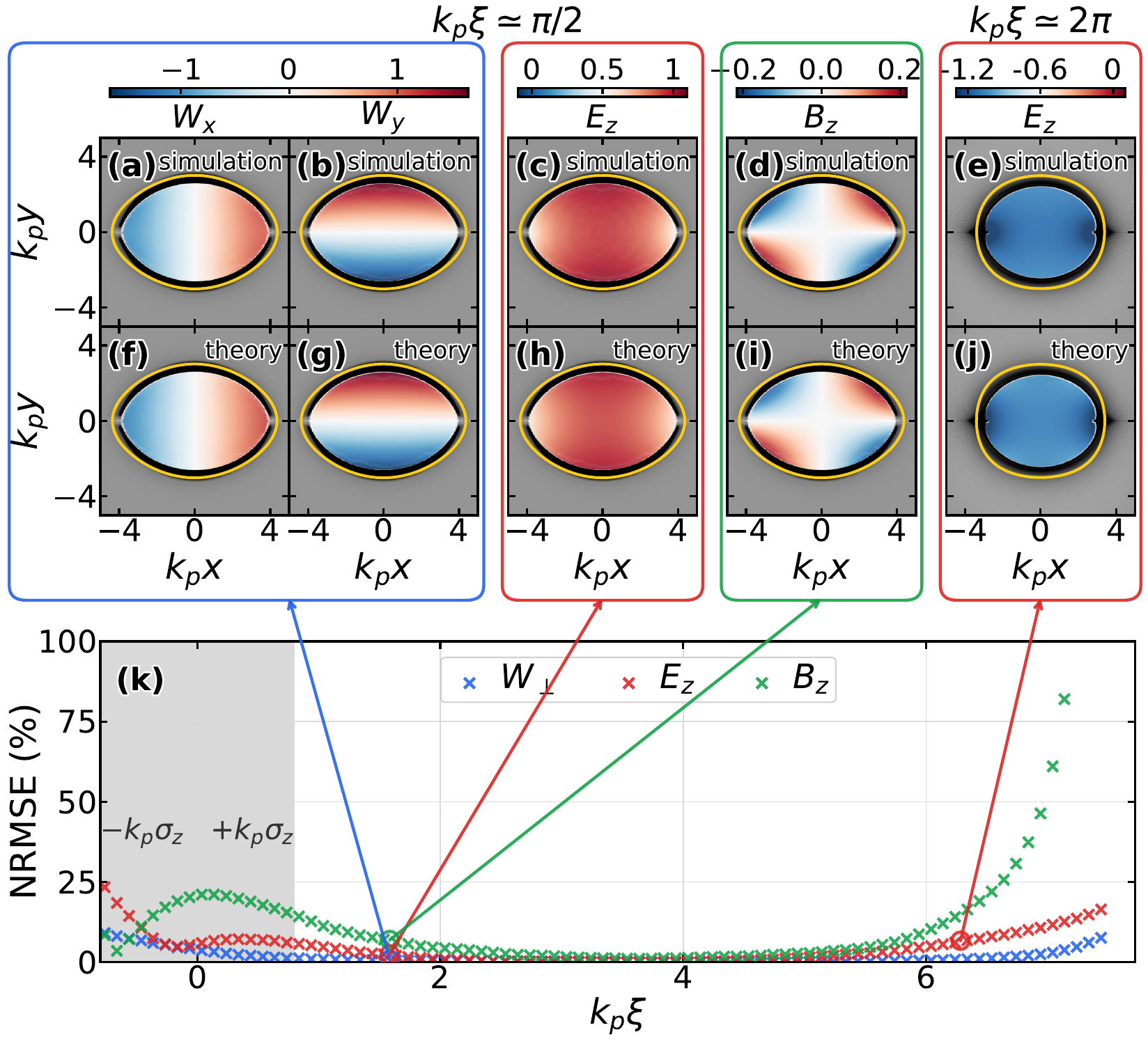}
    \caption{Comparison of simulated and theoretical wakefield maps for the same driver as Fig.~\ref{fig:asymmetry}, following the same conventions. (a)--(e) PIC results and (f)--(j) theoretical predictions from Eq.~\eqref{eq:general_moment_equations}, showing \(W_x\), \(W_y\), \(E_z\), and \(B_z\) at \(\xi\approx\pi/2\), and \(E_z\) at \(\xi\approx2\pi\); the inferred effective boundary \(R_b(\theta)\) is overlaid in yellow. (k) Normalized root-mean-square errors (NRMSEs) of \(\mathbf W_\perp\), \(E_z\), and \(B_z\) along \(\xi\), normalized to their corresponding RMS values inside the cavity; arrows connect the circled points to the corresponding field maps, with box colors matching the residual curves.}
    \label{fig:field_maps}
\end{figure}

To validate the model, we performed three-dimensional particle-in-cell (PIC) simulations with WarpX~\cite{Vay2018WarpX} for drivers spanning a range of peak densities and transverse aspect ratios. From the simulated fields we fit the wake moments \(\psi_k(\xi)\) inside the ion column and thereby infer the zero-potential boundary \(R_b(\theta)\). The wake-moment extraction procedure and the existence and uniqueness of the corresponding \(\delta\)-sheath boundary are detailed in Ref.~\cite{Yunbo_AAC_Paper_2026}.

Figure~\ref{fig:moments} compares the simulation-inferred wake moments with those obtained by numerically integrating Eq.~\eqref{eq:general_moment_equations}. Following the conventional maximum-radius initialization of axisymmetric blowout-boundary models~\cite{Tzoufras2008,Dalichaouch2021}, we launch Eq.~\eqref{eq:general_moment_equations} from the slice \(\xi_m\) where \(\psi_0'(\xi_m)=0\), near the region of maximum cavity extent. The simulation-inferred state at \(\xi_m\) provides the initial data for integration with no additional adjustable parameters. The theory then follows the longitudinal evolution of \(\psi_k(\xi)\) accurately, including the weak octupole moment \(\psi_4\). The accuracy in the wake moments translates directly into accurate predictions of both the transverse focusing fields and the longitudinal accelerating field, from Eqs.~\eqref{eq:focusing_field} and~\eqref{eq:longitudinal_field}.

Figure~\ref{fig:field_maps} demonstrates this agreement directly through comparisons of the theoretical and simulated field maps for the same case as Fig.~\ref{fig:asymmetry}, together with the corresponding normalized root-mean-square errors (NRMSEs) evaluated at each longitudinal slice along \(\xi\). The NRMSEs in Fig.~\ref{fig:field_maps}(k) remain at the percent level over most of the cavity, rising only within the driver region, where the wake is being excited and the blowout cavity is not yet completely formed, and near the rear of the blowout, where the sheath structure becomes increasingly complex. Including the next moment, \(k=6\), reduces the residuals further, especially at the rear of the blowout. Away from the head and tail, the octupole correction is relatively weak and the quadrupole moment dominates the asymmetric structure: the focusing fields \(W_x=-\partial_x\psi\) and \(W_y=-\partial_y\psi\) are predominantly linear but unequal in the two transverse directions. The accompanying transverse nonuniformity of \(E_z\) and the nonzero \(B_z\), which vanishes identically for axisymmetric blowout, are generated by the same asymmetric moment dynamics and therefore provide additional signatures of nonaxisymmetric structure. This advances the elliptical flat-beam analysis of Ref.~\cite{Manwani2025} in two respects: the cavity boundary is no longer restricted to an ellipse, and the asymmetry of \(W_x\) and \(W_y\) together with the nonuniformity of \(E_z\) follow self-consistently from the evolution of the wake moments rather than from an assumed cavity shape. 

We conclude the Letter by discussing the limitations, potential applications, and future work. The predictive domain of the present theory is set by the $\delta$-sheath model. As a mathematical idealization to reduce the total Lagrangian, the $\delta$-sheath accurately represents the interior wake potential and the focusing and accelerating fields inside the ion column, where the witness beam dynamics is usually determined. If the region near the sheath or outside the blowout boundary \(\partial\Omega_b\) is of interest, the present model needs to be extended to a finite-thickness sheath description, to capture the dynamics of the electron sheath and the region beyond it. Moreover, the theory can be extended to laser wakefield acceleration by introducing the laser-electron interaction Lagrangian \(L_{laser} =-\frac12\oint_{\partial\Omega_b} \left\langle a^2\right\rangle\sigma_S\,d\Gamma\), where $\langle a^2\rangle$ is the time-averaged square of the dimensionless laser amplitude. Further investigation and simulation verification are needed. 

For full predictive use without simulation-inferred initial data, the ultra-relativistic initialization provides a practical closure. The excitation phase is approximated by neglecting the electron Lagrangian \(L_e\) relative to the dominant electromagnetic contribution \(L_{\rm EM}\), following the same method used in Refs.~\cite{Golovanov2021,Golovanov2023}. Although this approximation does not constitute a complete theory of blowout formation, it correctly predicts the quadrupole moment \(\psi_2(\xi)\), which directly determines the asymmetry of the linear focusing channel. This is sufficient for simulation-free predictions of the longitudinally varying focusing strengths relevant to applications including asymmetric beam matching~\cite{Manwani2024}, emittance-mixing constraints~\cite{Diederichs2024}, and adiabatic plasma-lens design~\cite{Chen1990,Floettmann2014}. It also provides the field model needed to investigate the control of hosing instability in nonaxisymmetric focusing channels~\cite{Mehrling2017,Mehrling2019, Hildebrand2018}. A more complete excitation theory, including the transition into the proper blowout regime and the initialization of higher asymmetric moments, remains an important direction for future work.

This work was performed with the support of the US Dept. of Energy under Contract Nos. DE-SC0009914 and DE-SC0017648. This work used resources of the National Energy Research Scientific Computing Center, a US DOE Office of Science User Facility, operated under Contract No. DE-AC02-05CH11231.

\appendix
\section{Appendix}
\renewcommand{\theequation}{A\arabic{equation}}
\renewcommand{\thefigure}{A\arabic{figure}}
\setcounter{equation}{0}
\setcounter{figure}{0}
Using \(n_e(1-\bar v_z)\,d^2x_\perp=\sigma_S\,d\Gamma\) and applying \(\bar{\gamma}_e-\bar{p}_z=1+\psi\) on the boundary, the electron Lagrangian integral can be rewritten as
\begin{equation}
\begin{aligned}
L_e &= \int d^2x_\perp\,\mathcal L_e \\
&=
\int d^2x_\perp\,n_e(1-\bar v_z)
\frac{
\overline{1-\gamma_e^{-1}}
+(1-\bar v_z)\phi
+\bar v_z\psi
-\bar{\mathbf v}_\perp\cdot\mathbf A_\perp
}{
1-\bar v_z
} \\
&= \int_{\partial\Omega_b}d\Gamma\,\sigma_S\left[\frac12|\bm r_\perp'|^2+\phi-\bm r_\perp'\cdot\bm A_\perp\right].
\end{aligned}
\end{equation}
From Eq.~\eqref{eq:identity-delta-poisson}, the \(\phi\)-dependent part of the total Lagrangian reduces to zero because the integrand is proportional to \(\phi\):
\begin{equation}
\begin{aligned}
L_\phi &= \int_{\Omega_b}d^2x_\perp\,\nabla_\perp\phi\cdot\nabla_\perp\psi-\int_{\Omega_b}d^2x_\perp\,\phi+\int_{\partial\Omega_b}d\Gamma\,\sigma_S\phi \\
&= \int_{\partial\Omega_b}d\Gamma\,\phi\,(\partial_n\psi_{\rm in} + \sigma_S)-\int_{\Omega_b}d^2x_\perp\,\phi\,(\nabla_\perp^2\psi + 1)\\
&= 0 .
\end{aligned}
\end{equation}
The \(\bm A_\perp\)-dependent part before reduction is
\begin{equation}
\begin{aligned}
L_{\bm A_\perp}
&= \int_{\Omega_b}d^2x_\perp\left(\partial_\xi\bm A_\perp\cdot\nabla_\perp\psi-\frac12B_z^2\right)\\
&-\int_{\partial\Omega_b}d\Gamma\,(A_nV_n+A_\tau V_\tau)\sigma_S .
\end{aligned}
\end{equation}
The first term is integrated by parts in \(\xi\) using Reynolds' formula,
\(\frac{d}{d\xi}\int_{\Omega_b}d^2x_\perp\,\bm A_\perp\cdot\nabla_\perp\psi = \int_{\Omega_b}d^2x_\perp\,\partial_\xi\bm A_\perp\cdot\nabla_\perp\psi+\int_{\Omega_b}d^2x_\perp\,\bm A_\perp\cdot\partial_\xi\nabla_\perp\psi+\int_{\partial\Omega_b}d\Gamma\,V_n\bm A_\perp\cdot\nabla_\perp\psi\), because the transverse domain \(\Omega_b(\xi)\) moves with normal velocity \(V_n\). Dropping the resulting total \(\xi\)-derivative, we obtain
\begin{equation}
\begin{aligned}
L_{\bm A_\perp}
&= -\int_{\Omega_b}d^2x_\perp\,\bm A_\perp\cdot\partial_\xi\nabla_\perp\psi\\
&-\frac12\int_{\Omega_b}d^2x_\perp\,B_z^2-\int_{\partial\Omega_b}d\Gamma\,A_\tau V_\tau\sigma_S .
\end{aligned}
\end{equation}
Imposing Eq.~\eqref{eq:identity-delta-current} gives
\begin{equation}
\begin{aligned}
L_{\bm A_\perp}
&= \frac12\int_{\Omega_b}d^2x_\perp\,B_z^2-\int_{\partial\Omega_b}d\Gamma\,A_\tau\left(B_z+V_\tau\sigma_S\right)\\
&=\frac12\int_{\Omega_b}d^2x_\perp\,B_z^2 .
\end{aligned}
\end{equation}
Therefore, we obtain the reduced form in Eq.~\eqref{eq:reduced-lagrangian}.

Using Eq.~\eqref{eq:even_multipole_expansion}, the reduced Lagrangian can be rewritten in terms of the generalized coordinates given by the wake multipole moments \(\{\psi_k(\xi)\}\). The electromagnetic kinetic term is
\begin{equation}
\begin{aligned}
K_{\rm EM}
&=
\frac12\int_{\Omega_b}\left(E_z^2+B_z^2\right)d^2x_\perp
\\
&=
\frac12
\sum_{k,\ell}
\psi_k'\psi_\ell'
\int_0^{2\pi}\int_0^{R_b(\theta)}
r^{k+\ell+1}
\cos[(k-\ell)\theta]\,dr\,d\theta
\\
&=
\frac12
\sum_{k,\ell}
A^{\rm EM}_{k\ell}\psi_k'\psi_\ell',
\\[0.5ex]
A^{\rm EM}_{k\ell}
&=
\frac{1}{k+\ell+2}
\int_0^{2\pi}
R_b^{k+\ell+2}(\theta)
\cos[(k-\ell)\theta]\,d\theta .
\end{aligned}
\end{equation}
With \(d\Gamma=\left|\partial_\theta\bm R_b\right|d\theta=\sqrt{R_b^2+(\partial_\theta R_b)^2}\,d\theta\) and \(\partial_n\psi=\frac{\sqrt{R_b^2+(\partial_\theta R_b)^2}}{R_b}
\left.\partial_r\psi\right|_{r=R_b}\), the $\delta$-sheath kinetic term is
\begin{equation}
\begin{aligned}
K_e
&=
\frac12\oint_{\partial\Omega_b}
\left(V_n^2+V_\tau^2\right)\sigma_S\,d\Gamma
\\
&=
\frac12\oint_{\partial\Omega_b}
\frac{
\left(\left.\partial_\xi\psi\right|_{r=R_b}\right)^2
+
\left(\left.B_z\right|_{r=R_b}\right)^2
}{\sigma_S}\,d\Gamma\\
&=
-\frac12
\sum_{k,\ell}
\psi_k'\psi_\ell'
\int_0^{2\pi}
\frac{R_b^{k+\ell+1}(\theta)}
{\left.\partial_r\psi\right|_{r=R_b}}
\cos[(k-\ell)\theta]\,d\theta
\\
&=
\frac12
\sum_{k,\ell}
A^s_{k\ell}\psi_k'\psi_\ell',
\\[0.5ex]
A^s_{k\ell}
&=
-\int_0^{2\pi}
\frac{R_b^{k+\ell+1}(\theta)}
{\left.\partial_r\psi\right|_{r=R_b}}
\cos[(k-\ell)\theta]\,d\theta .
\end{aligned}
\end{equation}
The total kinetic term can be written in the same form with the coefficients \(A_{k\ell} = A^{\rm EM}_{k\ell} + A^s_{k\ell}\). Exploiting the symmetry \(A_{k\ell}=A_{\ell k}\), we write the kinetic contribution to the Euler--Lagrange equation as
\begin{equation}
\begin{aligned}
&\frac{d}{d\xi}\left(\frac{\partial K_{\rm tot}}{\partial\psi_k'}\right) - \frac{\partial K_{\rm tot}}{\partial\psi_k} \\
&=\sum_{\ell}A_{k\ell}\psi_\ell'' + \frac12\sum_{\ell,m}\left(\partial_{\psi_\ell}A_{km}+\partial_{\psi_m}A_{k\ell}-\partial_{\psi_k}A_{\ell m}\right)\psi_\ell'\psi_m',
\end{aligned}
\end{equation}
which gives the \(A_{k\ell}\) and \(B_{k\ell m}\) terms in Eq.~\eqref{eq:general_moment_equations}.

The electromagnetic potential term is
\begin{equation}
\begin{aligned}
U_{\rm EM} &= \frac12\int_{\Omega_b}|\nabla_\perp\psi|^2\,d^2x_\perp = \frac12\int_{\Omega_b}\psi\,d^2x_\perp \\
&= \frac12\int_0^{2\pi}\int_0^{R_b(\theta)}\left[-\frac{r^2}{4}+\sum_{k}\psi_k\,r^k\,\cos(k\theta)\right]r\,dr\,d\theta \\
&= \sum_{k}C_k\psi_k -\frac{1}{32}\int_0^{2\pi}R_b^4(\theta)\,d\theta, \\
C_k &= \frac{1}{2(k+2)}\int_0^{2\pi}R_b^{k+2}(\theta)\cos(k\theta)\,d\theta .
\end{aligned}
\end{equation}
The driver coupling term is
\begin{equation}
\begin{aligned}
L_b &= -\int_{\Omega_b}\rho_b(\xi,x,y)\psi(x,y)\,d^2x_\perp \\
&= \sum_{k}\lambda_k\psi_k + \frac14\int_0^{2\pi}\int_0^{R_b(\theta)} \rho_b r^3\,dr\,d\theta, \\
\lambda_k &= -\int_0^{2\pi}\int_0^{R_b(\theta)} \rho_b\!\left(\xi,r, \theta\right)r^{k+1}\cos(k\theta)\,dr\,d\theta .
\end{aligned}
\end{equation}
When evaluating potential and driver derivatives, the moving-boundary terms vanish, \(\int_0^{2\pi}R_b f(R_b,\theta,\xi)\psi(R_b,\theta;\{\psi_j\})\partial_{\psi_k}R_b\,d\theta = 0\), because the zero-potential boundary condition \(\psi(R_b,\theta;\{\psi_j\})=0\) is imposed. Therefore, only the bulk variations contribute to the \(C_k\) and \(\lambda_k\) terms in Eq.~\eqref{eq:general_moment_equations}.

\bibliography{references}

\end{document}